\documentstyle[stwol]{article} 

\def\Journal#1#2#3#4{{#1} {\bf #2}, #3 (#4)}

\def\NCA{\em Nuovo Cimento}
\def\NIM{\em Nucl. Instrum. Methods}
\def\NIMA{{\em Nucl. Instrum. Methods} A}
\def\NPB{{\em Nucl. Phys.} B}
\def\PLB{{\em Phys. Lett.}  B}
\def\PRL{\em Phys. Rev. Lett.}
\def\PRD{{\em Phys. Rev.} D}
\def\ZPC{{\em Z. Phys.} C}

\def\be{\begin{equation}}
\def\ee{\end{equation}}
\def\bea{\begin{eqnarray}}
\def\eea{\end{eqnarray}}

\begin{document}
\title{QUANTUM DEFORMATIONS OF SPACE-TIME SYMMETRIES WITH MASS-LIKE
DEFORMATION PARAMETER}

\author{ J. LUKIERSKI}
\address{Institute for Theoretical Physics,
University of Wroc\l{}aw,
Pl.\ Maxa Borna 9, 50-204 Wroc\l{}aw, Poland}

\date{}

\def\X{{\cal X}}
\def\M{{\cal M}}
\def\k{{\kappa}}
\def\kdef{$\k$-deformed }
\def\kdn{$\k$-deformation}
\def\poin{Poincar\'e }
\def\d{\delta}
\def\D{\Delta}
\def\e{\epsilon}
\def\h{\hat}
\def\ben{\begin{enumerate}}
\def\een{\end{enumerate}}
\def\P{{\cal P}}
\def\gpa{\h \P_{3,1}^{(g_{\mu\nu})}}
\def\U{{\cal U}}
\def\Uk{{\cal U}_\kappa}
\def\hPk{\h \P_\k}
\def\bel{\begin{equation}\label}
\def\ee{\end{equation}}
\def\tens{\otimes}
\def\r#1{(\ref{#1})}
\def\i{\item}
\def\kmin{$\kappa$-Minkowski }
\def\beq{\begin{eqnarray}}
\def\eeq{\end{eqnarray}}
\def\lbl{\label}
\def\pok#1{\frac{P_0}{#1\kappa}}
\def\pok{\frac{P_0}{\kappa}}
\def\cop{\triangle}
\def\triv#1{#1\tens 1 + 1 \tens #1}
\def\void{\,\,\cdot\,\,}
\def\ba{\begin{array}}
\def\ea{\end{array}}
\def\Lam{\Lambda}
\def\ost{\frac1{\sqrt2}}
\def\sh{\sinh}
\newcounter{rown}
\def\bl{\setcounter{rown}{\value{equation}}
        \stepcounter{rown}\setcounter{equation}0
        \def\theequation{\arabic{rown}\alph{equation}}
	}
\def\el{\setcounter{equation}{\value{rown}}
        \def\theequation{\arabic{equation}}
	}
\def\theequation{\thesection.\arabic{equation}}
\def\bl{\setcounter{rown}{\value{equation}}
        \stepcounter{rown}\setcounter{equation}0
        \def\theequation{\thesection.\arabic{rown}\alph{equation}}
	}
\def\el{\setcounter{equation}{\value{rown}}
        \def\theequation{\thesection.\arabic{equation}}
	}
\def\sec{\setcounter{equation}0}
\def\p{\partial}
\def\derp#1#2{\frac{\p #1}{\p #2}}
\def\u#1{\underline{#1}}
\def\hx{\hat x}
\def\hp{\hat p}
\def\lam{\lambda}
\def\h{\hat}
\def\({\left(}
\def\){\right)}
\def\<{\left<}
\def\>{\right>}
\def\[{\left[}
\def\]{\right]}







\twocolumn[\maketitle\abstracts{
The difficulties with the measurability of classical space-time distances are 
considered. We outline the framework of quantum deformations of
$D=4$ space-time symmetries with dimensionfull deformation parameter,
and present some recent results.}]
\sec
\section{Introduction}

It is well-known that two fundamental constants --- light velocity
$c$ and Planck constant $\hbar$ --- are introduced respectively by 
relativistic kinematics (Einstein's special relativity) and quantum 
mechanics.
In quantum mechanics the noncommutativity of the position and momentum 
observables 
\bel{1.1}
[\hat x_i,\,\hat p_j] = i\hbar \delta_{ij}\,,
\ee
implies the uncertainty relation 
\bel{1.2}
\D_\phi \hx_i \D_\phi \hp_i \stackrel{df}{=}\D x\,\D p \geq \frac{\hbar}2\,,
\ee
where $x^\phi_i = \< \phi| \hx_i | \phi \>$ is a mean position and
$
\D_\phi \hx_i = \(\< \phi |(\hx_i - x^\phi_i)^2 |\phi \>\)^{\frac12}\,.
$
In standard quantum mechanics the commutativity of the position operators 
$\hx_i$ implies the possibility to measure the position of quantum particle 
with arbitrary accuracy. Due to this property the Schr{\H o}dinger wave 
function $\psi(\vec x,t)$ is a classical field, with the arguments described 
by commuting space-time coordinates.

Recently there has been a considerable progress in the description of 
noncommutative or "quantum" geometry, which deals with algebra of functions 
on a "noncommutative manifold". The simplest example is provided by quantum 
phase space \r{1.1} and the algebra of functions $f(\hx,\hp)$.
However, one can consider the noncommutative structure also in space-time 
--- by assuming that
$[\hx_\mu,\,\hx_\nu]\neq 0$ ($\mu,\nu=0,1,2,3$) (see e.g.\ [1]). 
Physically, nonvanishing 
commutation relations of
space-time coordinates could be the effects caused by quantum gravity 
(see e.g.\ [2-5]) or quantum string theory (see e.g.\ [6-9]).
Below we shall outline some of the arguments.

\subsection{Elementary Planck length and quantum gravity}

It is known that quantum mechanics (Heisenberg uncertainty relation \r{1.2})
and relativistic kinamatics put together allows to consider the concept of
particle only in the space intervals larger that the Compton wave lenght
(for simplicity we drop the three-space vector indices):
\bel{1.4}
\D x > \frac\hbar{m_0 c}\,.
\ee
Indeed, because for relativistic particles energy 
$E=c(p^2+m_0^2c^2)^{\frac12}$ we  
have $\D E = c \D p \frac{p}{(p^2+m_0^2c^2)^{\frac12}}$ and for $p 
>\!\!> m_0^2$ 
one can write $\D x \D E \sim c \D x \D p \geq \hbar c$. If $\D 
E$ is larger or equal to the rest energy $m_0c^2$ the concept of mass looses
its
meaning [10]. We see therefore that if we put $\D E_{m} < m_0c^2$, one
gets \r{1.4} from \r{1.2}.

The uncertainty relation \r{1.4} leads effectively to the
existence of fundamental length, where $m_0$ is the rest
mass of the stable particle, if the creation and anihilation processes would not
take place. Because for $E>>m_0 c^2$ this is not the case,
therefore one should look
for the universal limitations on $\D x$ from below in another place e.g.\ 
in gravity theory, describing the space-time manifold as a dynamical system. 
The advantage of gravity is its universal nature, its coupling to any matter 
in the universe.

Let us consider the measurement process of the length in general relativity. 
Let us observe that the Einstein equations for the metric 
\bel{1.6}
\partial^2 g \sim \frac1{\k^2} \rho\,,
\ee
imply the following relation between the fluctuations of the metric $\D g$ 
and the fluctuation of the energy density $\rho = \frac{\D E}{ (\D x)^3}$
\bel{1.7}
\frac{\D g}{(\D x)^2} \sim \frac1{\k^2} \frac{\D E}{(\D x)^3}\,.
\ee
Because photon localizing with accuracy $\D x$ should have energy larger
than  $ E=\hbar \nu = \frac{\hbar}{\D x}$, one gets
\bel{1.8}
(\D g)(\D x)^2 > \frac\hbar{\k^2} = \lam^2_p\,.
\ee
Writing $(\D s)^2 = g (\D x)^2 > (\D g) (\D x)^2$ one gets $\D s > 
\lam_p$ where 
\bel{1.9}
\lam_p \simeq 1.6\cdot 10^{-33} \mbox{ cm}
\ee
is the Planck length.

The impossibility of localizing in quantized general relativity an event 
with the accuracy below the Planck length follows from the creation of 
gravitational field by the energy necessary for the measurement process. 
 
It should be mentioned that similar conclusion can be reached in the 
framework of the lattice quantum gravity and functional integraction 
approach to quantum gravity [11,12].

\subsection{Elementary length  and string theory}

The string theories (or rather superstrings theories which do not have 
tachyons and have consistent string loop expansions  --- see e.g.\ [13]) 
introduce the fundamental string length $\lam_s$ by the dimensionfull string 
tension $T$ 
as follows
\bel{1.10}
\lam_s^2 = \frac{h}{\pi T} = \frac{\hbar^2}{M_s^2}\,,
\ee
where $M_s$ denotes the fundamental string mass [6]. The Regge slope 
$\alpha'$ of the string trajectories is given by the formula 
\bel{1.11}
\alpha'= \frac{1}{2\pi T}\,.
\ee
The relation between the string mass and the Planck mass $M_p$ is obtained 
from the description of the graviton-graviton scattering by the string tree 
amplitude, with dimensionless string coupling constant $g$. One obtains that 
the Newton constant $G=\lam_p^2$ is given by 
\bel{1.12}
G=g^2 \frac1{M_s^2}
\ee
i.e\ one obtains $M_s= g M_p$. Indeed, the quantum gravity perturbative 
series in the coupling constant $\frac1{M^2_{p}}$ correspond to the string 
perturbative expansion with the coupling constant $\frac{g^2}{M_s^2}$.

The quantum mechanical uncertainty relation \r{1.2} is 
modified for the fundamental strings. The uncertainty in the position $\D x$ 
is the sum of two terms:
\ben
\i[a)] standard term is due to Heisenberg uncertainty relation for 
point-like canonically quantized objects,
\i[b)] new term is related with the size of the string increasing linearly 
with 
the energy.
\een 

One obtains (see e.g.\ [8,9])
\bel{1.13}
\D x \geq \frac{\hbar}{\D p} + k l_p^2 \D p\,.
\ee
The minimal value of $\D x $ is obtained for $(\D p)^2 \simeq \frac 
{\hbar}{k l_p^2}$ i.e.
\bel{1.14}
\D x \geq 2 \sqrt{k\hbar} l_p\,.
\ee

We see therefore that again it follows from the fundamental string theory 
that the Planck fundamental length  $l_p$ describes the accuracy of the 
measurements of space-time distances.

\sec
\section{Quantum \kdn s and noncommutative space-time}

In order to introduce the noncommutative space-time coordinates one could
follow two approaches:
\ben
\i[i)] One can consider the noncommutative Minkowski space only as the
representation space of the \poin symmetries, without identification of
the space-time coordinates with the translation sector of the \poin group.
In such a case the algebraic properties of the space-time coordinates and
the translation generators belonging to the \poin group might be
different; in particular one can introduce the \u{noncommutative}
Minkowski space coordinates and \u{classical} \poin symmetries
with commuting \poin group parameters. Such an
approach, outside of the framework of quantum groups was recently proposed
by Doplicher, Fredenhagen and Roberts [14].

\i[ii)] In another approach we introduce the noncommutative Minkowski
space described by the translation sector of the quantum \poin group. 
In such an approach the quantum \poin group 
as well as the quantum \poin algebra are the examples of noncommutative
and noncocommutative Hopf algebras, which provide the algebraic
generalization of the notions of the Lie group as well as the Lie algebra. 
 First quantum deformation
$\Uk(\P_{4})$ of $D=4$ \poin algebra has been proposed in [15], with
mass-like quantum deformation parameter $\k$. The introduction of
fundamental mass parameter $\k$ does not modify the nonrelativistic $O(3)$
symmetries and one gets the following noncommuting \kdef Minkowski space
coordinates (see e.g. [16-18])
\bel{2.1}
[\hx_{i},\hx_{j}]=0,\qquad [\hx_{i},\hx_{0}]=\frac{i}{\k} \hx_{i}\,,
\ee
\een
It is easy to see from \r{2.1} that for standard \kdef \poin symmetries the
space directions are classical and the quantum deformation affects the
time direction. In particular the mass shell condition is modified as
follows:
\bel{3.1}
p_{0}^{2}-\vec p^{2} = m^{2} \to (2\k\sinh\frac{p_{0}}{2\k})^{2} - \vec
p^{2}=m^{2}\,,
\ee
where the fourmomentum coordinates $p_{\mu}$ are the commuting variables.
One can introduce corresponding \kdef free KG fields and
the space-time picture in two different ways:
\ben
\i[i)] One can introduce the commuting space-time $x_{\mu}$ coordinates by
standard Fourier transform. In such a case the deformed free K-G equation 
takes the form
\bel{3.2}
[\D-(2\k\sinh\frac{\p_{t}}{2\k})^{2}-m^{2}]\varphi)x=0\,.
\ee
Such a deformation of scalar field theory was firstly considered in [19].
\i[ii)] Recently there were introduced fields $\varphi(\hat x)$ depending
on the noncommutative Min\-kow\-ski space coordinates [20-22]. In such a case
after introducing the noncommutative differential calculus on \kdef
Minkowski space \r{2.1} the \kdef Klein-Gordon equation 
takes the classical form:
\bel{3.3}
[(\derp{}{\hat x_{i}})^{2}-(\derp{}{\hat x_{0}})^{2}-m^{2}] \phi (\hat x) =0
\ee

\een

\sec
\section{Generalized \kdn s of $D=4$ relativistic symmetries}

The most general
class of noncommutative space-time coordinates 
described by the translation sector of quantum \poin group was
considered in [23]. One gets the following algebraic relations:
\bel{2.2}
 (R-1)_{\mu\nu}{}^{\rho\tau}(\hx_\rho \hx_\tau + \frac1\k
T_{\rho\tau}{}^\lam \hx_\lam+\frac1{\k^2} C_{\rho\tau})=0\,,
\ee
where the matrix $R$ describes the quantum $R$-matrix for the
Lorentz group satisfying the condition $R^2=1$, $\k$ is a masslike 
deformation parameter and $T_{\mu\nu}{}^\rho$,
$C_{\mu\nu}$ are the numerical coefficients (for details see [23]) which
are dimensionless. The condition $R^2$=1 can be removed if we consider
quantum \poin groups belonging to larger class of so called braided Hopf
algebras (see e.g.\ [24]). 

The relations \r{2.1}
follows as a special case of the relation \r{2.2}, with $R=\tau$ 
($\tau(a\tens b)=b\tens a$)  describing classical Lorentz symmetry, 
$C_{\mu\nu}=0$ and a particular choice of $T_{\mu\nu}{}^\rho$.

Recently there were also 
considered the \kdn s along one of the space axes, for
example $x^{3}$ (see [25]; this is so called tachyonic \kdn\ with
$O(2,1)$ classical subalgebra). Other interesting \kdn\ is the null-plane
quantum \poin algebra [26] with the ``quantized'' light  cone coordinate 
$x_{+}=x_{0}+x_{3}$ and classical $E(2)$ subalgebra. The generalized
\kdn s of $D=4$ \poin symmetries were recently proposed in [27] and describe
the \kdn\ in any direction $y_{0}=R^{0}{}^{\mu}x_{\mu}$ in Minkowski
space. Because the change of the linear basis in standard Minkowski space
$x_{\mu} \to y_{\mu}=R_{\mu}{}^{\nu} x_{\nu}$ implies the replacement of
Minkowski metric 
$
\eta^{\mu\nu} \to g^{\mu\nu} = R^{\mu}{}_{\rho}\eta^{\rho\tau}
R_{\tau}{}^{\nu}\,,$  ($R^{\mu}{}_{\rho}=(R_{}{\rho}{}^{\mu})^{T}$)
generalized \kdn s are obtained by deforming the classical \poin algebra with
arbitrary symmetric metric $g^{\mu\nu}$.  From
the formulae presented in [27] follows distinction between the \kdn s in the case
$g_{00}\neq0$ and $g_{00}=0$. It appers that 
\ben
\i[i)] If $g_{00}\neq0$ the \kdn\ is described by the classical $r$-matrix
satisfying  modified Y-B equation. Further, it can be shown [28] that the
minimal dimension of the bicovariant differential calculus on \kdef
Minkowski space is five, i.e.\ in the equation \r{3.3} one obtains
\bel{3.6}
d\varphi(\hat x) = d \hat x_{\mu}\derp{}{\hat x_{\mu}}\phi + \omega 
\Omega \phi
\ee
where e.g.\ for standard
\kdn\ described by \r{2.1} the additional one-form 
$\omega=d(x^{2}+\frac{3i}{\kappa}x^{0})-2x_{\mu}dx^{\mu}$  and $\Omega$ 
is the  nonpolynomial vector field described by \kdef mass Casimir. 
When $g_{00}\neq0$ the basic relations of \kdef differential calculus are the
following [20]
\bl
\beq\lbl{3.7a}
[d \hx_{\mu}, \hx_\nu ]&=&\ba[t]{l}\frac i \k g^0{}_\mu d\hx_\nu \\
                  +\frac i \k g_{\mu\nu} d\hx_0+\frac14g_{\mu\nu} \omega\,,
\ea \\
\lbl{3.7b}[\omega,\hx_\mu]&=&-\frac{4}{\k^2} d\hx_\mu\,.
\eeq
\el
\i[ii)]
If $g_{00}=0$ the \kdn\  is described by the classical $r$-matrix 
satisfying classical YB equation, which permits the extension of such \kdn\ 
to the conformal algebra [29]. Further, it can be shown that if $g_{00}=0$ 
the differential calculus is fourdimensional, with standard basis of the 
one-forms described by $d\hx_\mu$. The formula \r{3.6} has only first term on 
the rhs, and the commutator $[d\hx_\mu,\hx_\nu]$ can be obtained from 
\r{3.7a} by neglecting the last term on the rhs. 
\een

Having the differential calculi on \kdef Minkowski spaces one can consider 
the corresponding \kdef field theory for both cases $g_{00}\neq0$ 
and $g_{00}=0$.  This programme is now under 
consideration [22].

\def\Journal#1#2#3#4{{#1} {\bf #2}, #3 (#4)}

\def\NCA{\em Nuovo Cimento }
\def\NIM{\em Nucl. Instrum. Methods}
\def\NIMA{{\em Nucl. Instrum. Methods} A }
\def\NPB{{\em Nucl. Phys.} B }
\def\PLB{{\em Phys. Lett.}  B }
\def\PLA{{\em Phys. Lett.}  A }
\def\PRL{\em Phys. Rev. Lett. }
\def\PRD{{\em Phys. Rev.} D }
\def\ZPC{{\em Z. Phys.} C }
\def\IJMP{{\em Int. J. Mod. Phys. }}
\def\IJMPA{{\em Int. J. Mod. Phys.} A }
\def\JMP{{\em J. Mod. Phys. }}
\def\MPL{{\em Mod. Phys. Lett. }}
\def\MPLA{{\em Mod. Phys. Lett.} A }
\def\JPA{{\em J. Phys.} A }

\section*{References}
\def\b{\bibitem}

\end{document}